\documentclass[reqno,11pt]{article}

\usepackage{mathtools}
\usepackage{amsmath}
\usepackage{amsfonts}
\usepackage{amssymb}
\usepackage{graphicx}
\usepackage{epstopdf}

\usepackage[inner=2.5cm, outer=2.5cm, bottom=2.5cm, top=2.5cm]{geometry}

\newcommand*{\eqb}{\begin{equation}}
\newcommand*{\eqe}{\end{equation}}
\newcommand*{\al}{\alpha}

\newcommand{\norm}[1]{\left\| #1 \right\|}

\newcommand{\scalar}[1]{\left\langle #1 \right\rangle}

\newcommand{\R}{\mathbb{R}}

\newcommand{\pr}{\mathbf{P}}

\newcommand{\bol}[1]{\mathbf{#1}}

\begin{document}
\title{Explicit Densities of Multidimensional L\'evy Walks}

\makeatletter
    \@addtoreset{equation}{section} 
\makeatother

 \def\theequation{\arabic{section}.\arabic{equation}}

  \setcounter{page}{1}
  \thispagestyle{empty}

 \author{\normalsize Marcin Magdziarz $^1$, Tomasz Zorawik $^2$}
\date{}

\maketitle

\begin{abstract}
We provide explicit formulas for asymptotic densities of the 2- and 3-dimensional  ballistic L\'evy walks. It turns out that in the 3D case the densities are given by elementary functions. The densities of the 2D L\'evy walks are expressed in terms of hypergeometric functions  and the right-side Riemann-Liouville fractional derivative which allows to efficiently evaluate them numerically. The theoretical results  agree with Monte-Carlo simulations. The obtained functions solve certain differential equations with the fractional material derivative.
\end{abstract}

\bigskip
\bigskip
\bigskip
L\'evy walks play an important role in statistical physics. For the first time they were analyzed in \cite{shlesinger klafter, klafter blumen} in the framework of generalized master equations.  What distinguishes L\'evy walks is a strong spatial-time coupling. Particles move at a constant velocity $v_0$ between turning points and choose a new direction at random \cite{klafter book, cont_levy}.  The time between each turn has a slowly decaying power-law tail $\psi(t)\propto t^{-1-\al}$ with $0<\al<1$, but the motion is ballistic \cite{asymptotic densities}. These features cause that L\'evy walks find many practical applications including  bacteria swimming \cite{bacteria}, blinking nanocrystals \cite{blinking crystals}, light transport in optical materials \cite{light} and human travel \cite{human 1, human 2}. For more we refer to \cite{levy walks}, which is also a good introduction to the topic. 

In the resent article \cite{asymptotic densities} the shape of density profiles of 1-dimensional L\'evy walks in the asymptotic long-time limit was found. However  many  observed phenomena are 2- or 3-dimensional \cite{bacteria} and up till now there were no methods of determining the probability density function (PDF) in this scenario. In this paper we solve this problem and introduce methods to find the explicit formulas for asymptotic PDFs both in 2- and 3-dimensional case. The 2D result is given by Eq.~(\ref{cartesian}). The 3D result given by Eq.~(\ref{3d result}) has a particularly elegant and simple form.  Moreover these PDFs solve certain differential equations \cite{MSSZ} with the fractional material derivative \cite{material derivative 1, material derivative 2}. 

From mathematical point of view multidimensional L\'evy walk is defined  as \cite{MSSZ}:
\[
\bol{L}(t)= \sum_{i=1}^{N(t)}v_0T_i\bol{V}_i+\left(t-\sum_{i=1}^{N(t)}T_i\right)v_0\bol{V}_{N(t)+1},
\]
where $
N(t)=\max\{k\geq 0:\sum_{i=1}^k T_i \leq t\}$ is a renewal process, $T_i$ is a sequence of independent, identically distributed (IID) random variables with power-law distribution $\psi(t)\propto t^{-1-\al}$ governing time between turns and $\bol{V}_i$ is a sequence of IID random vectors governing a direction of the motion \cite{klafter book}. For simplicity we assume $v_0=1$.

In physical and mathematical literature appear also so-called \textit{wait-first L\'evy walks} and \textit{jump-first L\'evy walks} \cite{levy walks} defined as coupled continuous time random walks \cite{Montroll}. These processes have different properties, but the techniques presented here allow us to calculate the asymptotic PDF of these multidimensional processes as well. 

Levy walk  is rotationally invariant - each direction of the motion is equally possible. Therefore to determine the asymptotic PDF of $\bol{L}(t)$ it is enough to determine the asymptotic PDF of the radius $\norm{\bol{L}(t)}$. In this work we use two main ideas. The first one concerns determining the asymptotic PDF of $L_1(t)$ -  the projection of $\bol{L}(t)$ on the first axis. The second one is finding the relation between the distribution of $\norm{\bol{L}(t)}$ and $L_1(t)$.

\section{2D case}
After each turn the particles choose the direction $\bol{V}_i$  randomly. We assume that each $\bol{V}_i$ has a uniform distribution on a circle. Let $G(\bol{x},t)$ denote the PDF of $\bol{L}(t)$, $\bol{x}=(x_1,x_2)\in \R^2$. In  \cite{MSSZ} authors found the limit process $\bol{X}(t)$ such that $n^{-1}\bol{L}(nt)\stackrel{n\rightarrow \infty}\longrightarrow \bol{X}(t)$ in Skorokhod $J_1$ topology \cite{skorokhod}. Calculating the PDF $H(\bol{x},t)$ of the limit process $\bol{X}(t)$ is equivalent to finding the desired asymptotic behavior of $G(\bol{x},t)$, that is a function $H(\bol{x},t)$ such that $ G(\bol{x},t)\propto H(\bol{x},t)$. We know \cite{MSSZ} that in the  Fourier-Laplace transform  space $H$ is given by 
\eqb \nonumber
H(\bol{k},s)=\frac{1}{s}g\left(\frac{i\bol{k}}{s}\right)=\frac{1}{s}\frac{\int_{\bol{S}^1}\left(1-\scalar{\frac{i\bol{k}}{s}, \bol{u}}\right)^{\alpha-1} K(d\bol{u})}{\int_{\bol{S}^1}\left(1-\scalar{\frac{i\bol{k}}{s}, \bol{u}}\right)^\alpha K(d\bol{u})},
\eqe
where $\bol{k}=(k_1,k_2)\in\R^2$ is the Fourier space variable, $s$ is the Laplace space variable, $K(d\bol{u})$ is a uniform distribution on a circle  $\bol{S}^1$ and $ \scalar{\;,\;}$ denotes an inner product in $\R^2$. Now we can notice that
$
H_1(k_1,s)=H((k_1,0),s)
$
gives us the Fourier-Laplace transform of PDF of $X_1(t)$ - the projection of $\bol{X}(t)$ on the first axis. A marginal distribution of the uniform distribution on a circle $K_1(d\bol{u})$ has the density \cite{feller}
\eqb
K(du_1)=\frac{1}{\pi}\frac{1}{\left(1-u_1^2\right)^{1/2}} du_1,
\eqe
thus 
$
H_1(k_1,s)=\frac{1}{s}g_1\left(\frac{ik_1}{s}\right),
$
where
\eqb
g_1(\xi)=\frac{\int_{-1}^1(1-\xi u)^{\alpha-1}\frac{1}{\pi \left(1-u^2\right)^{1/2}}du}{\int_{-1}^1(1-\xi u)^\alpha\frac{1}{\pi \left(1-u^2\right)^{1/2}}du}.
\eqe
Since we are now dealing with a 1-dimensional process $X_1(t)$ we can use methods to invert the F-L transform from \cite{asymptotic densities} which apply only to such processes. This gives us
$
H_1(x,t)=\frac{1}{t}\Phi_1\left(\frac{x}{t}\right),
$
 where
\eqb
\Phi_1(x)=-\frac{1}{\pi}\lim_{\epsilon \rightarrow 0} \operatorname{Im}\left[\frac{1}{x+i\epsilon}g_1\left(-\frac{1}{x+i\epsilon}\right)\right].
\eqe
After transformations we obtain
\eqb
\label{2d_phir}
\Phi_1(x)=-\frac{1}{\pi x}\operatorname{Im}\frac{_{2}F_1((1 - \alpha)/2, 1 - \alpha/2; 1; \frac{1}{x^2})}{_2F_1(-\alpha/2, (1- \alpha)/2; 1; \frac{1}{x^2})}
\eqe
for $x\in(-1,1)$ and $\Phi_1(x)=0$ otherwise. Here $_2F_1(a,b;c;x)$ is a hypergeometrical function \cite{hiper} defined as
\eqb
_2F_1(a,b;c;x)=\sum_{k=0}^\infty \frac{(a)_k (b)_k}{(c)_k}\frac{x^k}{k!}
\eqe
for $|x|<1$ and analitacaly continued for $x>1$. In this series $(a)_k=\frac{\Gamma[a+k]}{\Gamma[a]}$ is the Pochhammer symbol. The hypergeometric functions can be evaluated numerically and are implemented in most of the mathematical packages including Matlab and Mathematica. Their use permits high-precision calculations.

\begin{figure}
\includegraphics[width=8.1cm]{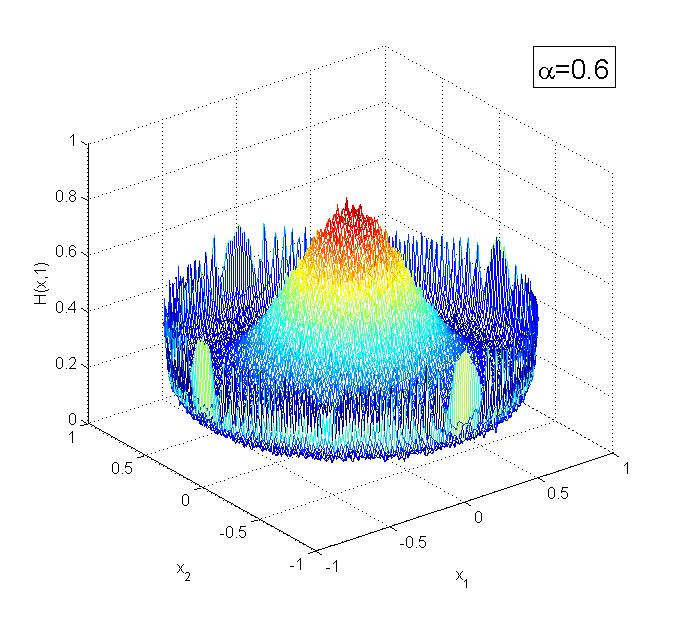}
\includegraphics[width=8.1cm]{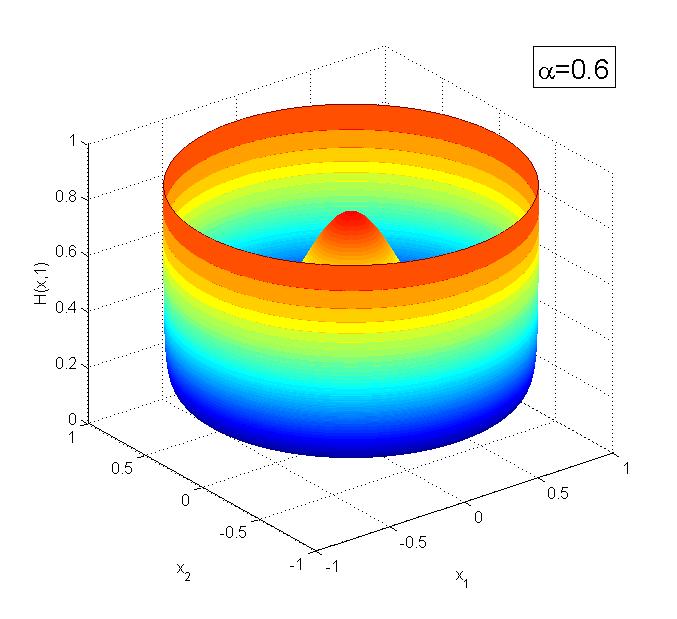}
\caption{\label{2d fig}Density $H(\bol{x},1)$ of 2D L\'evy walk with $\alpha0.6$ estimated using Monte Carlo methods from $10^7$ trajectories (left) and obtained from Eq.~(\ref{cartesian}) (right).}
\end{figure}

We now relate the PDF $H_R(r,t)$ of the radius $\norm{\bol{X}(t)}$ to the calculated the   PDF of $X_1(t)$. The rotational invariance of L\'evy walk $\bol{L}(t)$ implies the rotational invariance of the limit process $\bol{X}(t)$. Therefore to find $H(\bol{x},t)$ it suffices to determine $H_R(r,t)$. From  $H(\bol{k},s)=\frac{1}{s}g\left(\frac{i\bol{k}}{s}\right)$ we deduce the scaling  $H_R(r,t)=\frac{1}{t}\Phi_R\left(\frac{r}{t}\right)$. The factorisation of $\bol{X}(1)$ into radial
and directional parts gives us
\eqb
\bol{X}(1)\stackrel{d} =\norm{\bol{X(1)}}\bol{V},
\eqe
where $\bol{V}$ is a random vector uniformly distributed on a circle, indepedent of $\norm{\bol{X(1)}}$ and "$\stackrel{d}=$" denotes the equality of distribution. This implies 
\eqb
\label{2d relation}
\pr (|X_1(1)|\leq x)=\frac{2}{\pi}\int_0^1\frac{1}{(1-u^2)^{1/2}}\pr\left(\norm{\bol{X}(1)}\leq\frac{x}{u}\right)du
\eqe
for $x\geq0$. The differentiation of the above equation yields
\eqb
\Phi_1(x)=\frac{1}{\pi}\int_0^1\frac{1}{(1-u^2)^{1/2}}\frac{1}{u}\Phi_R\left(\frac{x}{u}\right)du
\eqe
and after some calculations we arrive at
\eqb
\label{2d-eq}
\Phi_R(r)=-2\pi^{1/2}rD_-^{1/2}\{\Phi_1(x^{1/2})\}(r^2),
\eqe
where  $D_-^{1/2}$ is the right-side Riemann-Liouville fractional derivative of order $1/2$ \cite{frac}:
\eqb
\label{rl_derivative}
D_-^{1/2}\{f(x)\}(y)=-\frac{d}{d y}\frac{1}{\pi^{1/2}}\int_y^\infty \frac{f(x)}{(x-y)^{1/2}}dx.
\eqe
This fractional derivative can be calculated numerically, for instance see \cite{matlab} for Matlab code. Combining Eqs.~(\ref{2d_phir}) and (\ref{2d-eq}) gives us
\begin{eqnarray}
&&\Phi_R(r)= \nonumber\\ 
&&\quad-\frac{2r}{\pi^{1/2}}D_-^{1/2}\Big\{\frac{1}{x^{1/2}}\operatorname{Im}\frac{_{2}F_1((1 - \alpha)/2, 1 - \alpha/2; 1; \frac{1}{x})}{_2F_1(-\alpha/2, (1- \alpha)/2; 1; \frac{1}{x})}\Big\}(r^2) \nonumber
\end{eqnarray}
for $r\in(0,1)$. In Cartesian coordinates $H(\bol{x},t)$ can be calculated as
\eqb
H(\bol{x},t)=\frac{1}{2\pi t\norm{\bol{x}}}\Phi_R\left(\frac{\norm{\bol{x}}}{t}\right).
\eqe
Therefore
\begin{eqnarray}
\label{cartesian}
&&H(\bol{x},t)=-\frac{1}{\pi^{3/2}t^2} \times \\
&&\quad D_-^{1/2}\Big\{\frac{1}{x^{1/2}}\operatorname{Im}\frac{_{2}F_1((1 - \alpha)/2, 1 - \alpha/2; 1; \frac{1}{x})}{_2F_1(-\alpha/2, (1- \alpha)/2; 1; \frac{1}{x})}\Big\}\left(\frac{\norm{\bol{x}}^2}{t^2}\right) \nonumber
\end{eqnarray}
for $\norm{\bol{x}}<t$ and $H(\bol{x},t)=0$ in the opposite case.
The right panel of FIG.~\ref{2d fig} presents $H(\bol{x},1)$ for $\alpha=0.6$ calculated from Eq.~(\ref{cartesian}). The result is in agreement with the density pictured in the left panel of FIG.~\ref{2d fig} estimated using Monte Carlo (MC) methods from $10^7$ trajectories. The algorithm for simulating L\'evy walks is presented in \cite{MT}. The drawback of using this algorithm is that it only gives the approximation of the asymptotic behavior of these processes. We observe that our Eq.~(\ref{cartesian}) gives much more accurate results than MC methods. The right panel of FIG.~\ref{2d fig} shows that $H(\bol{x},1)\rightarrow \infty$ when $\norm{\bol{x}}\rightarrow 1$ and MC methods reveal it only after a large number of simulations which is time-consuming.
\section{3D case}
\begin{figure}
\centering
\includegraphics[width=8.6cm]{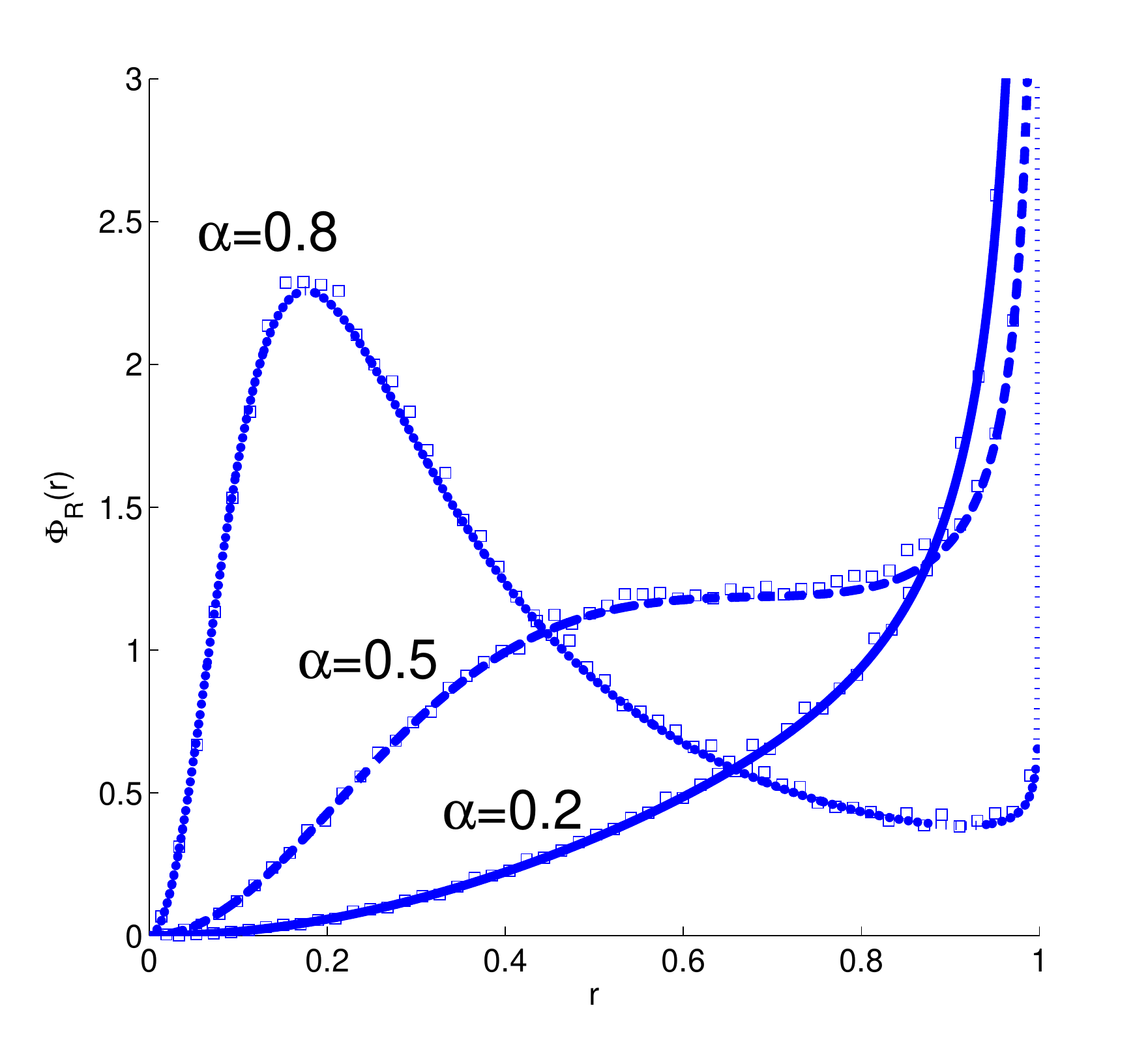}
\caption{\label{cont_3d}Density $\Phi_R(r)$ of 3D L\'evy walk obtained from Eq.~(\ref{3d result}) for $\alpha=0.2$, $\alpha=0.5$ and $\alpha=0.8$ (lines). Theoretical results are compared with densities estimated using Monte Carlo methods (squares).}
\end{figure}

For  3D process $\bol{X}(t)=(X_1(t), X_2(t), X_3(t))$ we use similar methods.  The notation remains the same. The random vectors $\bol{V}_i$ have a uniform distribution $K(d\bol{u})$ on a sphere $\bol{S}^2$. The Fourier-Laplace transform of $H(\bol{x},t)$ is (see \cite{MSSZ})
\eqb
H(\bol{k},s)=\frac{1}{s}\frac{\int_{\bol{S}^2}\left(1-\scalar{\frac{i\bol{k}}{s}, \bol{u}}\right)^{\alpha-1} K(d\bol{u})}{\int_{\bol{S}^2}\left(1-\scalar{\frac{i\bol{k}}{s}, \bol{u}}\right)^\alpha K(d\bol{u})},
\eqe
where $\bol{k}=(k_1,k_2,k_3)\in\R^3$ is the Fourier space variable, $s$ is the Laplace space variable  and $ \scalar{\;,\;}$ denotes an inner product in $\R^3$. In the 3D case applying the same reasoning as in the 2D case yields simpler results. This is caused by the fact that a one dimensional marginal distribution $K_1(du_1)$ of $K(d\bol{u})$ is uniform on the interval $[-1,1]$ (see \cite{feller}). Folowing the reasoning from the 2D case we obtain that the distribution of $X_1(1)$ is expressed by elementary functions
\begin{eqnarray}
&&\Phi_1(x)=\frac{2(\alpha+1)}{\pi\alpha}\sin(\pi\alpha)\times \nonumber\\
&&\quad\quad\quad\frac{(1-x^2)^{\alpha}}{(1-x)^{2\alpha+2}+(1+x)^{2\alpha+2}+2\cos(\pi\alpha)(1-x^2)^{\alpha+1}} \nonumber  
\end{eqnarray}
for $x\in (0,1)$, $\Phi_1(x)=\Phi_1(-x)$ for $x\in(-1,0)$ and $\Phi_1(x)=0$ otherwise. Equation~(\ref{2d relation}) now has the form 
\eqb
\pr (|X_1(1)|\leq x)=\int_0^1\pr\left(\norm{\bol{X}(1)}\leq \frac{x}{u}\right)du
\eqe
and the following relation between $\Phi_R$ and $\Phi_1$ holds:
\eqb
\label{3d-eq}
\Phi_R(r)=-2r\Phi_1'(r).
\eqe
Therefore
\begin{eqnarray}
&&\Phi_R(r)=\frac{8}{\pi}\frac{\al+1}{\al}\sin(\pi\al)r(1 - r^2)^{\al-1} \times\nonumber\\
&&\quad\quad\quad\quad\quad\frac{(1 + r)^{2+2\al}(1+\al - r) - (1-r)^{2+2\al}(1+\al + r)-2r(1-r^2)^{1+\al}\cos(\pi\al)}
{\left((1 + r)^{2+2\al} + (1 - r)^{2+2\al}+2(1-r^2)^{1+\al}\cos(\pi\al)\right)^2}
\label{3d result}
\end{eqnarray}
for $r\in(0,1)$ and $\Phi_R(r)=0$ in the opposite case. Figure~\ref{cont_3d} presents $\Phi_R(r)$ for different values of $\alpha$ calculated from Eq.~(\ref{3d result}). We notice that the theoretical results are in perfect agreement with the simulations. However contrary to using Eq.~(\ref{3d result}) the MC methods require a lot of time to get precise results.

Concluding, in this Letter we derived the exact formulas for the asymptotic densities of the 2- and 3-dimensional ballistic L\'evy walks. The result has a simple form, especially in the 3D case. The methods presented here can be successfully applied to different types of L\'evy walks operating in a ballistic regime. Furthermore, the same techniques allow us to calculate the PDF of $n$-dimensional L\'evy walks where the number of dimensions $n$ is arbitrary. We hope that these results will prove useful in practical applications.

\section*{Acknowledgement}
The authors thank Vasily Zaburdaev for interesting discussions.
\\
This research was partially supported by NCN Maestro grant no. 2012/06/A/ST1/00258.


 \bigskip \smallskip

 \it

 \noindent
$^1$ $^2$
Hugo Steinhaus Center, \\
Faculty of Pure and Applied Mathematics, \\
Wroclaw University of Technology, \\
Wyspianskiego 27, 50-370 Wroclaw, Poland. \\[4pt]
$^1$ e-mail: marcin.magdziarz@pwr.edu.pl\\
$^2$ e-mail: tomasz.zorawik@pwr.edu.pl

\end{document}